\documentclass[aps,prd,notitlepage,nofootinbib,superscriptaddress,longbibliography 
]{revtex4-1}

\usepackage{amsmath}
\usepackage{amssymb}
\usepackage{amsthm}
\usepackage{setspace}
\usepackage{color}
\usepackage[pdftex]{graphicx}
\long\def\symbolfootnote[#1]#2{\begingroup\def\thefootnote{\fnsymbol{footnote}}\footnote[#1]{#2}\endgroup}

\newcommand{\bra}[1]{\langle {#1} \vert}
\newcommand{\ket}[1]{\vert {#1} \rangle}

\begin{document}

\title{Maximal trace distance between isoenergetic bosonic Gaussian states}

\author{\firstname{T.J.}~\surname{Volkoff}}
\email{volkoff@konkuk.ac.kr}
\affiliation{Department of Physics, Konkuk University, Seoul 05029, Korea}%

\begin{abstract}
We locate the set of pairs $(\rho_{1},\rho_{2})$ of Gaussian states of a single mode electromagnetic field that exhibit maximal trace distance subject to the energy constraint $\langle a^{\dagger}a \rangle_{\rho_{1}}=\langle a^{\dagger}a \rangle_{\rho_{2}} = E$. Any such pair allows to achieve the minimum possible error in the task of binary distinguishability of two single mode, isoenergetic Gaussian quantum signals. In particular, we show that the logarithm of the minimal error probability for distinguishing two maximally trace distant, isoenergetic Gaussian states scales as $-E^{2}$, less than the achievable scaling of the minimal error probability for distinguishing, e.g., a pair of isoenergetic Heisenberg-Weyl coherent states with energy $E$ or a pair of isoenergetic quadrature squeezed states with energy $E$. For the case of a field consisting of $M>1$ modes, we locate the set of pairs of maximally trace distant isoenergetic, isocovariant Gaussian states. These results have basic applications in the theory of continuous variable quantum communications with Gaussian states of light.
\end{abstract}
\maketitle

\section{Introduction\label{sec:intro}}
Gaussian states (i.e., quasi-free states \cite{bratteli2}) on the canonical commutation relations (CCR) C$^{*}$ -
algebra generated by the Weyl operators of a single bosonic mode are fundamental in the theory of continuous variable quantum information processing due to the complete classification of their quantum dynamics \cite{holevoonemode}. In this sense, they have a similar standing in continuous variable quantum mechanics as states on $B(\mathbb{C}^{2})$, i.e., ``qubit'' states, have in finite-dimensional quantum mechanics. Several metrics on the space of trace class, positive operators on a separable Hilbert space $\mathcal{H}$ can be used to quantify the distance between two Gaussian quantum states \cite{morozova}. However, the distance which most directly quantifies the distinguishability of two quantum density operators $\rho_{1}$ and $\rho_{2}$ is the trace distance $\Vert \rho_{1} - \rho_{2} \Vert_{1}:= \text{tr} \vert \rho_{1} - \rho_{2} \vert$, i.e., the distance between $\rho_{1}$ and $\rho_{2}$ considered as elements of $B_{1}(\mathcal{H})$, the Schatten $p=1$ ideal of $B(\mathcal{H})$. Specifically, the trace distance appears in a central theorem of quantum estimation theory \cite{helstrombook}: given two normal, faithful states $\omega_{1}$ and $\omega_{2}$ of a von Neumann algebra with respective densities $\rho_{1}$ and $\rho_{2}$ present with equal \textit{a priori} probability, the minimal error probability $p_{\mathrm{err}}(\rho_{1},\rho_{2})$ that can be obtained in the task of distinguishing $\rho_{1}$ and $\rho_{2}$ by outcomes of any quantum measurement is given by 
\begin{equation}
p_{\mathrm{err}}(\rho,\sigma) :={1\over 2} - {1\over 4}\Vert \rho - \sigma \Vert_{1}.
\end{equation}
The simplicity of this theorem hides several difficulties. For instance, whereas a closed formula exists for, e.g., the Bures metric on the set of Gaussian quantum states \cite{pirandolafidel}, no analogous formula exists for the trace distance between two generic Gaussian states. Furthermore, the fact that any two Gaussian states $\rho_{1}$ and $\rho_{2}$ that have ``finite energy'' exhibit trace distance $\Vert \rho_{1} - \rho_{2} \Vert_{1}<2$, complicates the analysis of the quantum information theory for continuous variables. In the first part of this work, we take $\mathcal{H}$ to be the Hilbert space $\ell^{2}(\mathbb{C})$, defined as the norm closure of the complex linear span of vectors of the form $\sum_{j=1}^{\infty}c_{n}\ket{n}$, where $\sum_{j=1}^{\infty}\vert c_{n} \vert^{2} < \infty$, $c_{n}\in \mathbb{C}$, and $\lbrace \ket{n} \rbrace_{n=0,1,\ldots}$ is the orthonormal basis of Fock vectors. Furthermore, a density operator $\rho \in B_{1}(\mathcal{H})$ is said to have \textit{finite energy} if and only if the condition $\langle a^{\dagger}a \rangle_{\rho}:= \text{tr} a^{\dagger}a \rho < \infty$, where $a\ket{n}=\sqrt{n}\ket{n-1}$, $a^{\dagger}\ket{n}=\sqrt{n+1}\ket{n+1}$ are the annihilation and creation operators, respectively, is satisfied. Note that the finite energy states form the dense domain of the operator $a^{\dagger}a$ and that the linear functional $\rho \mapsto \text{tr}a^{\dagger}a\rho$ is lower semicontinuous on $B_{1}(\mathcal{H})$ \cite{holevoqubook}.

In the case of two finite energy, pure Gaussian states $\ket{\psi_{1}}$, $\ket{\psi_{2}}$, the condition  $\Vert \ket{\psi_{1}}\bra{\psi_{1}}-\ket{\psi_{2}}\bra{\psi_{2}}\Vert_{1}<2$ is equivalent to $\vert \langle \psi_{1} \vert \psi_{2} \rangle \vert > 0$, due to the formula
\begin{equation}
\Vert \ket{\psi_{1}}\bra{\psi_{1}} - \ket{\psi_{2}}\bra{\psi_{2}} \Vert_{1}=2\sqrt{1 - \vert \langle \psi_{1}\vert \psi_{2} \rangle \vert^{2}}.
\label{eqn:fidelrestate}
\end{equation} It then follows that for any isometry $V:\mathcal{H}\rightarrow \mathcal{K}$ between Hilbert spaces $\mathcal{H}$ and $\mathcal{K}$ with $\ket{\psi_{1}}$, $\ket{\psi_{2}} \in \text{dom}V$ (e.g., a noiseless quantum computation), the states $V\ket{\psi_{1}}$, $V\ket{\psi_{2}}$ cannot be distinguished with certainty.

In this work, we consider the problem of locating the pairs of Gaussian states $(\rho_{1},\rho_{2})$ of a single quantum harmonic oscillator that exhibit maximal trace distance $ \Vert \rho_{1} - \rho_{2} \Vert_{1}$ subject to the energy constraint $\langle a^{\dagger}a \rangle_{\rho_{1}} = \langle a^{\dagger}a \rangle_{\rho_{2}} = E $. Before delving into the details of the proof, we will show that the pairs of energy constrained Gaussian states that achieve the maximum trace distance are pure, i.e., the maximum is obtained on certain pairs of pure states $(\ket{\varphi_{1}},\ket{\varphi_{2}})$ that satisfy the constraint $\langle \varphi_{j} \vert a^{\dagger}a \vert \varphi_{j} \rangle = E$, $j=1,2$.  To set the notation, consider the single mode Heisenberg-Weyl Lie algebra defined by the self-adjoint operators $q,p,\mathbb{I}$ such that $[q,p]=i\mathbb{I}$, $[q,\mathbb{I}]=[p,\mathbb{I}]=0$. We write $R=(R_{1},R_{2})$ for the vector of quadratures, where $R_{1}=q$, $R_{2}=p$.  A single mode Gaussian state $\rho$ can be fully specified by its characteristic function, defined for $z\in \mathbb{R}^{2}$ by $\chi_{\rho}(z):=\mathrm{tr }\; e^{iR^{T}z}\rho =  e^{-{1\over 2}z^{T}\Sigma_{\rho}z +im_{\rho}^{T}z}$, where $m_{\rho}:=\langle R \rangle_{\rho} \in \mathbb{R}^{2}$ is the mean vector and $\Sigma_{\rho}$ is the covariance matrix, viz., a positive matrix in $GL(2,\mathbb{R})$ with entries given by $(\Sigma_{\rho})_{i,j} = \text{tr} \rho \left( R_{i}-(m_{\rho})_{i} \right) \circ \left( R_{j}-(m_{\rho})_{j} \right)$, where $A\circ B:= {1\over 2}\left( AB + BA \right)$ is the Jordan product. Therefore, a general single mode Gaussian state is parametrized by five real numbers: two positive numbers representing the diagonal entries of $\Sigma_{\rho}$, one real number specifying the off-diagonal entry of $\Sigma_{\rho}$, and two real numbers specifying $m_{\rho}$.

We now define \begin{equation}p_{\mathrm{min}}:= \min_{\ket{\psi_{1}}\bra{\psi_{1}} , \ket{\psi_{2}}\bra{\psi_{2}}  \in G(\mathcal{H})_{E} } p_{\mathrm{err}}(\ket{\psi_{1}}\bra{\psi_{1}} , \ket{\psi_{2}}\bra{\psi_{2}})\label{eqn:mindef}\end{equation} where $G(\mathcal{H})_{E}$ is the set of Gaussian states on the single mode Hilbert space $\mathcal{H}=\mathrm{span}_{\mathbb{C}}\lbrace \ket{0},\ket{1},\ldots \rbrace$ that satisfy $\langle a^{\dagger}a\rangle = E$.  It follows that
\begin{eqnarray}
p_{\mathrm{min}}&:=& \min_{\ket{\psi_{1}}\bra{\psi_{1}} , \ket{\psi_{2}}\bra{\psi_{2}}  \in G(\mathcal{H})_{E} } p_{\mathrm{err}}(\ket{\psi_{1}}\bra{\psi_{1}} , \ket{\psi_{2}}\bra{\psi_{2}}) \nonumber \\ &\ge &   \min_{\rho , \sigma \in G(\mathcal{H})_{E} }p_{\mathrm{err}}(\rho,\sigma) \nonumber \\ {} &\ge & \min_{\rho , \sigma \in CCH(G(\mathcal{H})_{E}) }p_{\mathrm{err}}(\rho,\sigma).
\label{eqn:gaussianminerror}
\end{eqnarray}
where $CCH(G(\mathcal{H})_{E})$ is the closed, convex hull of $G(\mathcal{H})_{E}$ and, for convenience, we denote the set of energy constrained Gaussian states by $G(\mathcal{H})_{E}$.
That equality holds in all inequalities appearing in Eq.(\ref{eqn:gaussianminerror}) can be shown by the following chain of standard arguments: 1) the set of quantum states $\sigma$ that satisfy $\langle a^{\dagger}a \rangle_{\sigma}\le E$ is compact \cite{holevoqubook} and contains the closed set $G(\mathcal{H})_{E}$, which is, therefore compact, 2)  $CCH(G(\mathcal{H})_{E})$ is compact \cite{infanalysisbook} and its extreme points are the pure states of $G(\mathcal{H})_{E}$, 3) the joint convexity of trace distance implies that it attains its extrema on pure states $CCH(G(\mathcal{H})_{E})$. Similar arguments have been used in quantum information theory to show, e.g., that the output entropy of a completely positive, trace preserving linear map on a finite dimensional, energy-constrained quantum state space is attained on a pure argument of the map \cite{mancini}.

In order to solve for $p_{\mathrm{min}}$ in Eq.(\ref{eqn:mindef}), we first note that a pure Gaussian state $\rho$ is characterized by the property of satisfying the Heisenberg uncertainty principle with equality, or, equivalently, by the property that $\Sigma_{\rho}$ can be brought to the matrix $\mathrm{diag}(1/2,1/2)$ by a symplectic transformation of $\mathbb{R}^{2}$ (with symplectic form $\Delta = i\sigma_{y}$ where $\sigma_{y}$ is the Pauli matrix $(\sigma_{y})_{1,1}=(\sigma_{y})_{2,2}=0$, $(\sigma_{y})_{1,2}=\overline{(\sigma_{y})_{2,1}}=-i$) \cite{holevoqubook}. Therefore, a pure single mode Gaussian state is specified by 4 real numbers. Due to the energy constraint (see also Eq.(\ref{eqn:enerconstraint}) below) the minimization problem in (\ref{eqn:mindef}) is defined on a manifold of dimension $6$ ($=4+4-2$). Reduction of the problem of minimization of $\vert \langle \psi_{1} \vert \psi_{2} \rangle \vert^{2}$ to the problem of minimization of a function of one real variable is the main task undertaken in the following.

A brief outline of the paper is as follows:

\section{Calculation of $p_{\mathrm{min}}$\label{sec:calcn}}
By the Euler decomposition of the symplectic group $Sp(2,\mathbb{R})$, and the metaplectic representation of $Sp(2,\mathbb{R})$ on the quantum state space of a single quantum oscillator \cite{arvind}, every pure single mode Gaussian state can be written as  $\ket{ (\alpha , z)} := D(\alpha)S(z)\ket{0}$, where $S(z):= e^{{1\over 2}\left( \overline{z} a^{2} - za^{\dagger 2} \right)}$ is the unitary squeezing operator, $D(\alpha):=e^{\alpha a^{\dagger} - \overline{\alpha}a }$ is the unitary displacement operator,  $\ket{0}$ is the Fock vacuum state, and $\alpha , z \in \mathbb{C}$. The energy constraint  $\langle a^{\dagger}a \rangle = E$ for a pure, single mode Gaussian state ${\ket{ (\alpha , z)}}$ is given explicitly in terms of the displacement parameter $\alpha$ and squeezing parameter $z$ by \cite{mandel}
\begin{equation}
\vert \alpha \vert^{2}+\sinh^{2}\vert z \vert = E.
\label{eqn:enerconstraint}
\end{equation} For a pair of Heisenberg-Weyl coherent states $\ket{\pm \alpha}:= \ket{(\pm \alpha ,0)}$ that exhibit the lowest fidelity among energy constrained coherent states (i.e., $\vert \alpha \vert = \sqrt{E}$),  one obtains linear scaling of the logarithm of the Helstrom error probability, $-\log p_{\mathrm{err}} \sim \mathcal{O}(E) $. Logarithmic scaling of $-\log p_{\mathrm{min}}$ with the energy $E$ is obtained by considering centered Gaussian states $S(x)\ket{0}$ and $S(-x)\ket{0}$ ($x \in \mathbb{R}$) that are squeezed in the canonically conjugate quadratures $q$ and $p$, respectively. In particular, $\vert \langle 0 \vert S(-x)^{\dagger}S(x)\vert 0 \rangle \vert^{2} = \vert \langle 0 \vert S(2x)\vert 0 \rangle \vert^{2}=1/(2E+1)$.  We now show in Lemma 1 that the possibility of investing part of the allotted energy into quadrature squeezing allows to achieve $-\log p_{\mathrm{min}} \sim \mathcal{O}(E^{2})$.

\textbf{Lemma 1.} \textit{Let $p_{\mathrm{min}}$ denote the minimal error probability for distinguishing  $\ket{\varphi_{1}}$, $\ket{\varphi_{2}}\in G(\mathcal{H})_{E}$, present with equal} a priori \textit{probability. Then $p_{\mathrm{min}}$ satisfies }\begin{equation} -\log p_{\mathrm{min}} \ge 4E^{2}+\mathcal{O}(E). \label{eqn:errlowbd}\end{equation}

The minimal error probability $p_{\mathrm{err}}\left( \ket{\varphi_{1}},\ket{\varphi_{2}} \right)$ for distinguishing $\ket{\varphi_{1}}$ and $\ket{\varphi_{2}}$ (present with equal \textit{a priori} probability) with an arbitrary quantum measurement is given by \cite{helstrombook}: \begin{eqnarray}p_{\mathrm{err}}\left( \ket{\varphi_{1}},\ket{\varphi_{2}} \right) &=& {1\over 2}\left(1-\sqrt{1-\vert \langle \varphi_{1} \vert \varphi_{2} \rangle \vert^{2}}\right) \nonumber \\ &=& {1\over 4} \vert \langle \varphi_{1} \vert \varphi_{2}\rangle \vert^{2} + \mathcal{O}\left( \vert \langle \varphi_{1} \vert \varphi_{2}\rangle \vert^{4} \right) , \label{eqn:distingeq} \end{eqnarray} where the second line follows from the MacLaurin series of $\sqrt{1-x}$.  Therefore, the proof of Lemma 1 reduces to the task of identification of pure states $\ket{\varphi_{1}}$, $\ket{\varphi_{2}}$ such that $\vert \langle \varphi_{1} \vert \varphi_{2} \rangle \vert^{2}=e^{-4E^{2} + \mathcal{O}(E)}$, subject to the energy constraint. The proof is considerably simplified by noticing that the constrained minimization of the fidelity can be carried out over pairs of pure Gaussian states $\ket{ (\alpha_{1} , z_{1})}$, $\ket{ (\alpha_{2} , z_{2})}$ with real parameters $\alpha_{1} \le 0$, $\alpha_{2} \ge 0$, i.e., with mean vectors on the real axis and on opposite sides of the origin, a fact which we presently demonstrate.

\begin{figure}[t!]
\begin{center}
\includegraphics[scale=.5]{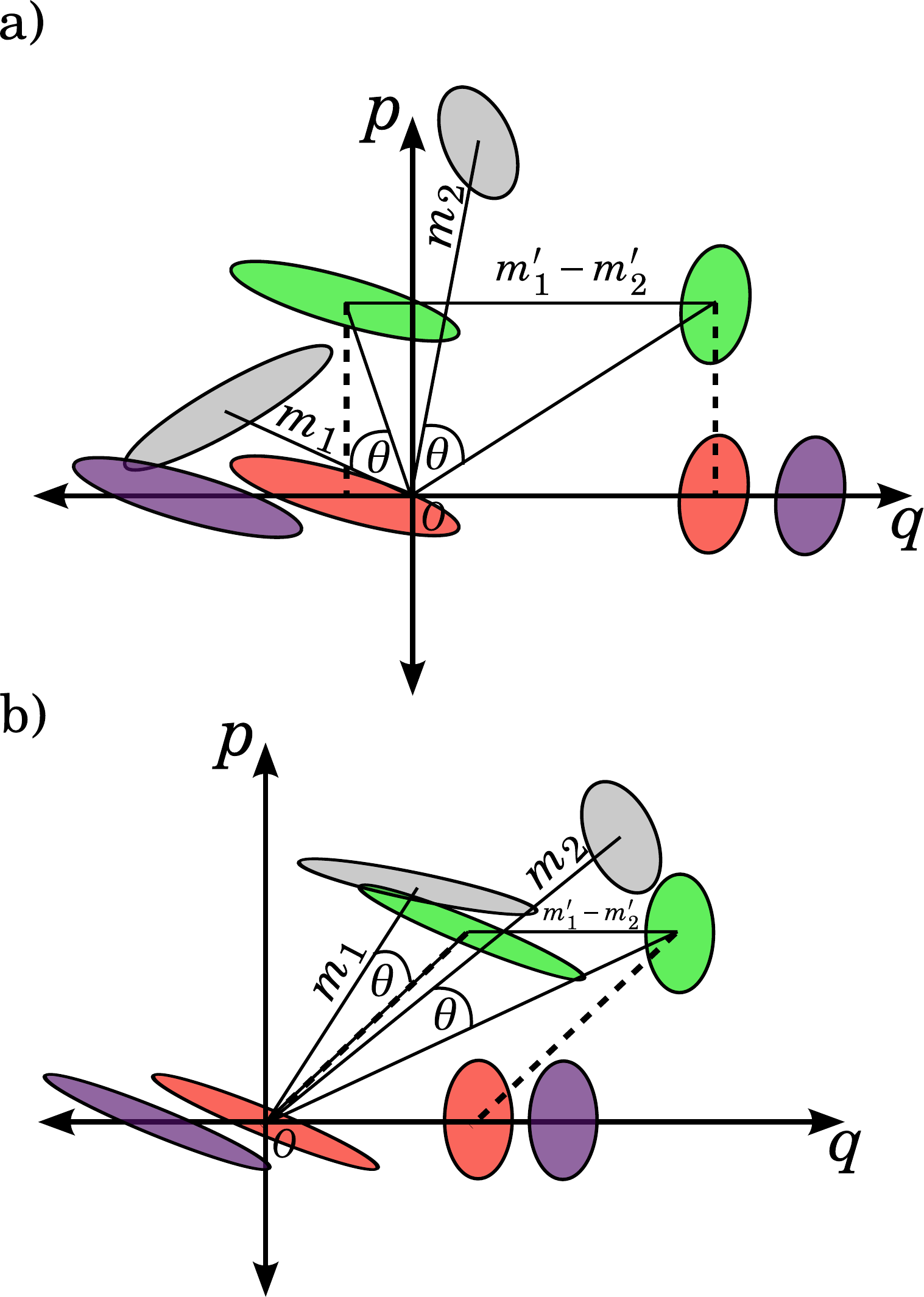}
\caption{\label{fig:ellip}Quadrature ellipses of the pure Gaussian states considered in the proof of Lemma 1. The major (minor) axis of each ellipse corresponds to the greatest (least) variance exhibited by a measurement of a quadrature. States corresponding to gray, green, and violet ellipses necessarily satisfy the energy constraint $\langle a^{\dagger}a \rangle = E$. Two cases are shown: a) $m_{1}'-m_{2}'$ parallel to $q$-axis when $m_{1}$ and $m_{2}$ are on opposite sides of the $p$-axis. b) $m_{1}'-m_{2}'$ parallel to $q$-axis when $m_{1}$ and $m_{2}$ are on the same side of the $p$-axis.}
\end{center}
\end{figure}

The fidelity $\vert \langle \varphi_{1} \vert \varphi_{2}\rangle \vert^{2}$ between two pure Gaussian states $\ket{\varphi_{1}}$ and $\ket{\varphi_{2}}$, with covariance matrices $\Sigma_{1}$, $\Sigma_{2}$, respectively, is given by \begin{equation}
\vert \langle \varphi_{1} \vert \varphi_{2}\rangle \vert^{2} = { e^{-{1\over 2}\left( m_{1}-m_{2} \right)^{T}\left( \Sigma_{1} + \Sigma_{2} \right)^{-1} \left( m_{1}-m_{2} \right)} \over \sqrt{ \det \left( \left( \Sigma_{1} + \Sigma_{2} \right) \right) }} ,
\label{eqn:purefidel}
\end{equation}
which can be obtained from a more general formula for the quantum affinity $\mathrm{tr}\sqrt{\rho_{1}}\sqrt{\rho_{2}}$ between quasifree states $\rho_{1}$ and $\rho_{2}$ \cite{holevoquasiequiv} or for the Bures distance $d_{B}(\rho_{1},\rho_{2}):=\mathrm{tr}\sqrt{ \sqrt{\rho_{1}}\rho_{2}\sqrt{\rho_{1}}} = \Vert \sqrt{\rho_{1}}\sqrt{\rho_{2}}\Vert_{1}$ between displaced, squeezed thermal states $\rho_{1}$, $\rho_{2}$ \cite{scutaru}. Consider the states $\ket{ (\alpha_{1} , z_{1})}$ and $\ket{ (\alpha_{2} , z_{2})}$, which are associated with mean vectors $m_{1(2)}=[\sqrt{2}\mathrm{Re}\alpha_{1(2)},\sqrt{2}\mathrm{Im}\alpha_{1(2)}]^{T}$ in $\mathbb{R}^{2}$ (Fig.(\ref{fig:ellip}a), gray ellipses). For an appropriate value of $\theta \in [0,\pi]$, the rotation $e^{-i\theta a^{\dagger} a}$ achieves the mapping $m_{1} -m_{2} \mapsto m_{1}'-m_{2}'$, where $m_{1}'-m_{2}'$ is parallel to the $q$-axis. The rotation preserves both the energy of each state and their inner product, producing the green ellipses in Fig.(\ref{fig:ellip}a). Projection of the mean vectors $m_{1}$, $m_{2}$ to the $q$-axis preserves the fidelity and lowers the expected energy of each state (Fig.(\ref{fig:ellip}a), red ellipses). In order for both of the resulting states to satisfy the energy constraint, the excess energy can be used to increase the magnitude of $m_{1}-m_{2}$, and thereby reduce the fidelity (Fig.(\ref{fig:ellip}a), violet ellipses). If the rotation by $e^{-i\theta a^{\dagger} a}$ produces a relative mean vector $m_{1}'-m_{2}'$ such that $m_{1}'$ and $m_{2}'$ are on the same side of the $p$-axis, or such that either of $m_{1}'$, $m_{2}'$ lie on the $p$-axis, the mean vector nearer the origin can be displaced to the origin, see, e.g., Fig.(\ref{fig:ellip}b), reducing the energy of both states. The excess energy is again used to scale $m_{1}-m_{2}$.

The proof of Lemma 1 now consists simply of choosing appropriate squeezing parameters $z_{j}$ defining the covariance matrices in Eq.(\ref{eqn:purefidel}).

\textbf{Proof of Lemma 1.}  We specialize to $\ket{\varphi_{1}} = D(r_{1})S(z_{1})\ket{0}$ and $\ket{\varphi_{2}}=D(r_{2})S(z_{2})\ket{0}$, with $r_{1}\le 0$, $r_{2}\ge 0$, and $z_{j}\ge 0$. By defining $d_{j} := e^{2z_{j}}$, the formula of Eq.(\ref{eqn:purefidel}) becomes $\vert \langle \varphi_{1} \vert \varphi_{2}\rangle \vert^{2} = F(r_{1},r_{2},d_{1},d_{2})$, where
\begin{equation}
F(r_{1},r_{2},d_{1},d_{2}) : = {2\sqrt{d_{1}d_{2}}\over d_{1} + d_{2}}e^{-{2d_{1}d_{2}\over d_{1} + d_{2}}\left( r_{2}-r_{1} \right)^{2} } .
\label{eqn:qsqueezedfidel}
\end{equation} Note that taking $z_{j}\ge 0$, i.e., squeezing of each state along the $q$ quadrature, corresponds to $d_{j}= 2(E-r_{j}^{2})+1 + 2\sqrt{(E-r_{j}^{2})^{2}+(E-r_{j}^{2})} $ due to the energy constraint.

For $d_{1}=d_{2}=d$, the function $F$ takes the simple form $F(r_{1},r_{2},d)=e^{-d(r_{2}-r_{1})^{2}}$. However, if $d_{1}=d_{2}$, the energy constraint, combined with the conditions $r_{1}\le 0$, $r_{2}\ge 0$, requires that $r_{1}=-r_{2} = -\sqrt{E-{1\over 4}\left( {d-1\over \sqrt{d}} \right)^{2}}$. The fidelity therefore simplifies to \begin{equation}
F(d)=e^{d^{2}-(4E+2)d+1}
\label{eqn:eqfidel}
\end{equation}
with parametric dependence on the energy $E$. On the interval $d\in [1,\infty)$, the minimum of the exponent of $F(d)$ is achieved for $d=d_{c}(E)$, where $d_{c}(E)=2E+1$. The corresponding value for the minimal fidelity, $F(d=d_{c}(E))$, is:
\begin{equation}
\vert \langle \varphi_{1} \vert \varphi_{2}\rangle \vert^{2} = e^{-4 E^{2} - 4E}.
\label{eqn:eqfidel2}
\end{equation}
where $\ket{\varphi_{1}}=\ket{  \left( - r(d_{c}(E)) , {1\over 2}\ln d_{c}(E) \right)}$, $\ket{\varphi_{2}}=\ket{  \left(  r(d_{c}(E)) , {1\over 2}\ln d_{c}(E) \right)}$ and \begin{eqnarray} r(d_{c}(E))&=&\sqrt{E-{1\over 4}\left( {d_{c}(E)-1\over \sqrt{d_{c}(E)}}\right)^{2}} \nonumber \\ &=& \sqrt{{E^{2}+E \over 2E+1}}. \label{eqn:pair} \end{eqnarray} Using Eq.(\ref{eqn:eqfidel2}) in Eq.(\ref{eqn:distingeq}) and collecting multiplicative prefactors into the exponential proves Lemma 1.
$\square$

In the limit of large $E$, the ratio $d_{c}(E)/r(d_{c}(E))^{2}$ approaches $4$. It is intriguing that the value of the squeezing parameter $d_{c}(E)$ obtained in the proof above coincides with the maximal squeezing parameter for which a squeezed thermal state with thermal energy $E$ exhibits a positive $P$-distribution \cite{kimknight}. In view of Eqs.(\ref{eqn:distingeq}) and (\ref{eqn:eqfidel2}), the question remains whether $p_{\mathrm{min}}$ is equal to ${1\over 2}\left(1-\sqrt{1-e^{-4 E^{2} - 4E}}\right) $. If this is true, then the minimal value of $\vert \langle \varphi_{1} \vert \varphi_{2} \rangle \vert^{2}$ for isoenergetic Gaussian states $\ket{\varphi_{1}}$, $\ket{\varphi_{2}}$ is given by $\vert \langle  \left( - r(d_{c}(E)) , {1\over 2}\ln d_{c}(E) \right) \vert \left(  r(d_{c}(E)) , {1\over 2}\ln d_{c}(E) \right)\rangle \vert^{2}=e^{-4 E^{2} - 4E}$, where $r(d_{c}(E))$ is defined in Eq.(\ref{eqn:pair}). We return to this question after we consider the special case $m_{1}=-m_{2}$ for the mean vectors $m_{1}$ and $m_{2}$ of $\ket{\varphi_{1}} = D(r_{1})S(z_{1})\ket{0}$ and $\ket{\varphi_{2}}=D(r_{2})S(z_{2})\ket{0}$, respectively.

\textbf{Lemma 2.} \textit{Let $\mathcal{G} \subset G(\mathcal{H})_{E} \times G(\mathcal{H})_{E} $ denote the set of pairs of pure Gaussian states with expected energy $E$ and mean vectors that are opposite, i.e., $m_{1} =  - m_{2}$.
Then} \begin{equation}
\min_{( \ket{\varphi_{1}}, \ket{\varphi_{2}} ) \in \mathcal{G} } \vert  \langle \varphi_{1} \vert \varphi_{2} \rangle \vert^{2} = e^{-4E^{2}-4E}.
\label{eqn:prop2}
\end{equation}

\textbf{Proof.}
Without loss of generality, consider again $\ket{\varphi_{1}} = D(r_{1})S(z_{1})\ket{0}$ and $\ket{\varphi_{2}}=D(r_{2})S(z_{2})\ket{0}$, with $r_{1}\le 0$, $r_{2}\ge 0$. Due to the energy constraint, if $\Vert m_{1} \Vert = \Vert m_{2} \Vert$, then $\vert z_{1} \vert = \vert z_{2} \vert$.
The covariance matrix $\Sigma_{j}$ corresponding to $\ket{(r_{j},z_{j})}$ is given by
\begin{equation}
\Sigma_{j} = {1\over 2}\left( \begin{array}{ccc}
\cosh 2 \vert z_{j} \vert - \cos \theta_{j} \sinh 2\vert z_{j} \vert & -\sin \theta_{j}\sinh 2\vert z_{j} \vert  \\
-\sin \theta_{j}\sinh 2\vert z_{j} \vert & \cosh 2 \vert z_{j} \vert + \cos \theta_{j} \sinh 2\vert z_{j} \vert  \end{array} \right)
\label{eqn:covariances}
\end{equation}
where $\theta_{j}:=\text{Arg}z_{j}$. Taking derivatives with respect to $\theta_{j}$, one finds that the critical points of the function \begin{equation}-\log \vert \langle  \varphi_{1} \vert \varphi_{2} \rangle  \vert^{2} = {1\over 2}(m_{1}-m_{2})^{T}\left( \Sigma_{1}+\Sigma_{2} \right)^{-1}(m_{1}-m_{2}) + {1\over 2}\log \det \left( \Sigma_{1}+\Sigma_{2} \right) \label{eqn:exponentialargfidel}\end{equation} with respect to the angles $\theta_{1}$ and $\theta_{2}$ occur at the angle pairs $\left( \theta_{1},\theta_{2} \right) \in \lbrace (0,0),(\pi ,0),(0,\pi),(\pi,\pi) \rbrace$. Because the energy constraint is independent of $\theta_{1}$, $\theta_{2}$, it does not have to be taken into account to find the critical angles of the function in Eq.(\ref{eqn:exponentialargfidel}).  We will not further consider the critical point $(\pi,\pi)$ because it does not yield a maximum of Eq.(\ref{eqn:exponentialargfidel}). In particular, for $r_{1}\le 0$, $r_{2}\ge 0$, and $x_{1},x_{2}>0$
\begin{eqnarray}
\vert \langle 0 \vert S(-x_{1})^{\dagger}D(r_{1})^{\dagger}D(r_{2})S(-x_{2}) \vert 0 \rangle \vert^{2} &=& {2\sqrt{b_{1}b_{2}}\over b_{1} + b_{2}}e^{-{2\over b_{1} + b_{2}}(r_{1}-r_{2})^{2}} \nonumber \\ 
&> & {2\sqrt{b_{1}b_{2}}\over b_{1} + b_{2}}e^{-{2b_{1}b_{2}\over b_{1} + b_{2}}(r_{1}-r_{2})^{2}} \nonumber \\
&=& \vert \langle 0 \vert S(x_{1})^{\dagger}D(r_{1})^{\dagger}D(r_{2})S(x_{2}) \vert 0 \rangle \vert^{2} 
\end{eqnarray}
where $b_{j}:=e^{2x_{j}}$. The critical points $(\pi ,0),(0,\pi)$ correspond to the same physical scenario of a pair of pure Gaussian states $\ket{\varphi_{1}}$, $\ket{\varphi_{2}}$ with mean vectors $(\sqrt{2}r_{1} , 0)$ and $(\sqrt{2}r_{2} , 0)$, and with one state squeezed along the $p$ quadrature and other state squeezed along the $q$ quadrature. Therefore, we only need to compare the critical points $(0,0)$ and $(\pi , 0)$. For $x_{1}=x_{2}=x>0$ and $b:= e^{2x}$, so that the energy constraint implies $-r_{1}=r_{2}=\sqrt{E-{1\over 4}{(b-1)^{2}\over b}}$, the inner product corresponding to the $(\pi , 0 )$ case is given by
\begin{equation} \vert \langle 0 \vert S(-x)^{\dagger}D(r_{1})^{\dagger} D(r_{2})S(x) \vert 0 \rangle \vert^{2} = {2b\over 1+b^{2}}e^{-{8b \over 1+b^{2}}\left( E-{1\over 4}{(b-1)^{2}\over b} \right) } \label{eqn:pizerosame}\end{equation} where the allowed values of $b$ are in   $[1,2E+1 +2\sqrt{E^{2}+E}]$. The only extremum of the right hand side of Eq.(\ref{eqn:pizerosame}) that lies in the interval of valid $b$ occurs for $b=1$, which corresponds to the pair of opposite-phase coherent states $\ket{\varphi_{1}}=\ket{-\sqrt{E}}$, $\ket{\varphi_{2}}=\ket{\sqrt{E}}$. However, for all $E$, recall that Lemma 1 gives a pair of pure Gaussian states $\ket{\varphi_{1}}$, $\ket{\varphi_{2}}$ with $(\theta_{1},\theta_{2})=(0,0)$ that saturates Eq.(\ref{eqn:prop2}).  Because $e^{-4E^{2}-4E}< e^{-4E} = \vert \langle -\sqrt{E} \vert \sqrt{E} \rangle \vert^{2}$ for all $E>0$. Therefore, a minimum occurs for the critical angles $(0,0)$, specifically, for the states $\ket{\varphi_{1}}=\ket{  \left( - r(d_{c}(E)) , {1\over 2}\ln d_{c}(E) \right)}$, $\ket{\varphi_{2}}=\ket{  \left(  r(d_{c}(E)) , {1\over 2}\ln d_{c}(E) \right)}$. For $\Vert m_{1} \Vert = \Vert m_{2} \Vert$, this minimum is a global one because Eq.(\ref{eqn:eqfidel}) is strictly convex in $d$.  $\square$

Lemmas 1 and 2 are the main tools used to simplify the proof of Theorem 1, below, which shows that $p_{\mathrm{min}}$ is indeed equal to ${1\over 2}\left(1-\sqrt{1-e^{-4 E^{2} - 4E}}\right) $, or, equivalently, that $\min_{\ket{\varphi_{1}}, \ket{\varphi_{2}} \in G(\mathcal{H})_{E}}\vert  \langle \varphi_{1} \vert \varphi_{2} \rangle \vert^{2}=e^{-4E^{2}-4E}$ for all energies $E$. The proof of this theorem proceeds by the introduction of a new optimization problem with a looser energy constraint.

\begin{figure}
\begin{center}
\includegraphics[scale=.5]{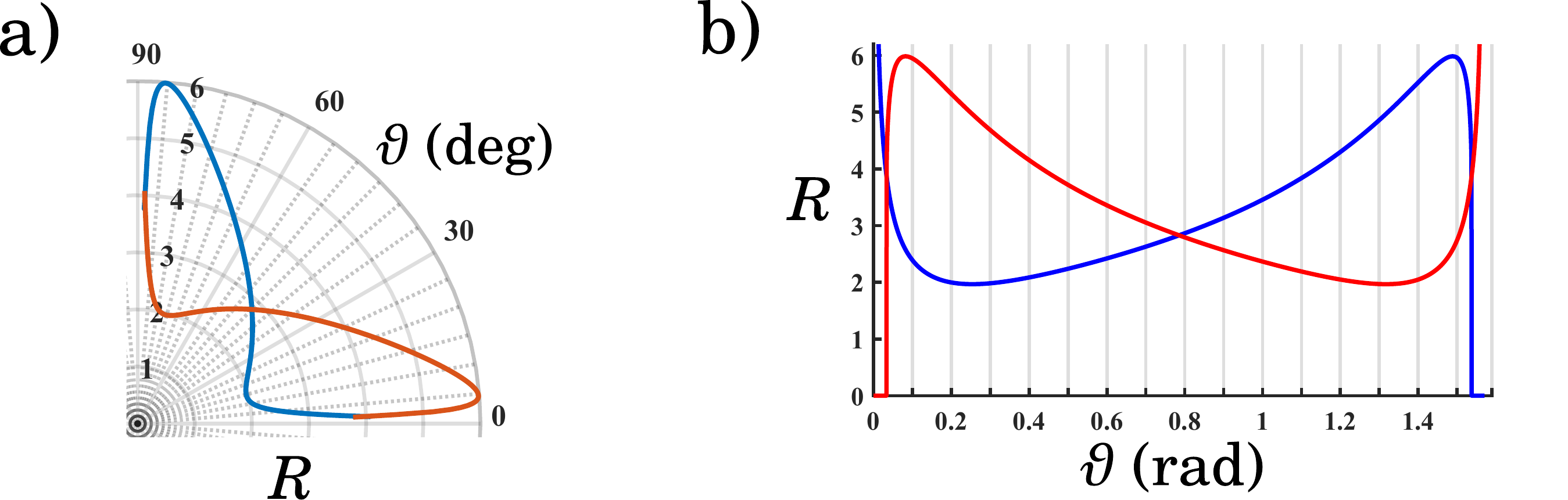}
\caption{\label{fig:polar}a) Polar plots of $R_{1}(\vartheta)$ (blue curve) and $R_{2}(\vartheta)$ (red curve) for $E=0.5$. b) Same curves as in a), shown relative to rectilinear axes.}
\end{center}
\end{figure} 

\textbf{Theorem 1.} $\mathrm{min}_{\ket{\varphi_{1}}, \ket{\varphi_{2}} \in G(\mathcal{H})_{E}}\vert  \langle \varphi_{1} \vert \varphi_{2} \rangle \vert^{2} = \mathrm{min}_{( \ket{\varphi_{1}}, \ket{\varphi_{2}} ) \in \mathcal{G} } \vert  \langle \varphi_{1} \vert \varphi_{2} \rangle \vert^{2} = e^{-4E^{2}-4E}$

\textbf{Proof.} Consider the less constrained problem (P1) below, which is defined by the constraint of total energy $2E$ to be distributed between two pure Gaussian states:

\underline{Problem (P1)}
\begin{eqnarray}
\ket{\varphi_{j}} \in G(\mathcal{H}) &\; , \;& j=1,2 \nonumber \\
 \vert \langle \varphi_{1} \vert \varphi_{2} \rangle \vert^{2} &\rightarrow& \mathrm{ min}, \nonumber \\
 \mathrm{tr}\left( a^{\dagger}a\sum_{i=1}^{2}\ket{\varphi_{j}}\bra{\varphi_{j}} \right)& = &2E \nonumber .
\end{eqnarray}
The same reasoning that led to Lemma 1 allows us to restrict the solution of (P1) to $\lbrace \ket{\varphi_{j}}\rbrace_{j=1,2}$ with mean vectors $(\sqrt{2}r_{j},0)$ such that $r_{1} \le 0$, $r_{2} \ge 0$. Furthermore, the unconstrained optimization over $\theta_{1}$, $\theta_{2}$ carried out in the proof of Lemma 2 is also valid here and allows us to restrict to $\ket{\varphi_{j}} = D(r_{j})S(w_{j})\ket{0}$ such that $w_{j}$ are real. In terms of $d_{j}:= e^{2w_{j}}$, the constraint in problem (P1) is given by
\begin{equation}
\sum_{i=1}^{2}r_{i}^{2}+{1\over 4}{(d_{i}-1)^{2}\over d_{i}} = 2E .
\end{equation}
Therefore, without loss of generality, one may take $r_{1} = -\sqrt{2E - r_{2}^{2} - f(d_{1},d_{2})}$, where $f(d_{1},d_{2}):=\sum_{i=1}^{2}{1\over 4}{(d_{i}-1)^{2}\over d_{i}} $. Eliminating $r_{1}$ from the expression for the fidelity results in
\begin{equation}
\vert \langle \varphi_{1} \vert \varphi_{2} \rangle \vert^{2} = {2\sqrt{d_{1}d_{2}}\over d_{1} + d_{2}}e^{-{2d_{1}d_{2}\over d_{1} + d_{2}}\left( \sqrt{2E - r_{2}^{2} -  f(d_{1},d_{2})} + r_{2} \right)^{2} },
\label{eqn:constr2}
\end{equation}
where $d_{j}>0$ and the only remaining constraint is that $2E - r_{2}^{2} -  f(d_{1},d_{2}) \ge 0$. A minimum of Eq.(\ref{eqn:constr2}) exists because the remaining constraint defines a compact subset of $\mathbb{R}^{3}$. At a local minimum of Eq.(\ref{eqn:constr2}), we must have ${d\over dr_{2}}\vert \langle \varphi_{1} \vert \varphi_{2} \rangle \vert^{2} = 0$. Taking the derivative, one finds that $r_{2}  = \mp \sqrt{ {2E - f(d_{1},d_{2}) \over 2} } = \pm r_{1} $. The case $r_{2}=r_{1}$ is not consistent with the fact that the minimum must occur for $r_{2}\ge 0$, unless $r_{2}=r_{1}=0$. For $r_{1}=r_{2}=0$, $\vert \langle \varphi_{1} \vert \varphi_{2} \rangle \vert^{2} > e^{-4E^{2}-4E}$ for all $E>0$ (see Appendix), so we conclude that the minimum must occur for the case of antipodal mean vectors, viz., $r_{2}=-r_{1}>0$. In this case, Eq.(\ref{eqn:constr2}) becomes 
\begin{equation}
\vert \langle \varphi_{1} \vert \varphi_{2} \rangle \vert^{2} = {2\sqrt{d_{1}d_{2}}\over d_{1} + d_{2}}e^{-{8d_{1}d_{2}\over d_{1} + d_{2}}\left( E - {1\over 2} f(d_{1},d_{2}) \right) } .
\label{eqn:constr3}
\end{equation}
The gradient of (\ref{eqn:constr3}) vanishes if and only if $g(d_{1},d_{2})=0$ and $g(d_{2},d_{1})=0$, where $g(d_{1},d_{2})$ is given by the following quartic:
\begin{eqnarray}
g(d_{1},d_{2}):= 2\left( d_{1}d_{2}^{3}+d_{1}^{3}d_{2}\right) + d_{2}^{2}-d_{1}^{2}-d_{1}d_{2}^{2}\left(16E + 8 \right) + 4d_{1}^{2}d_{2}^{2}.
\end{eqnarray}
Going to polar coordinates via $d_{1}=R\cos \vartheta$, $d_{2}=R\sin \vartheta$ transforms the problem of finding solutions to the system $g(d_{1},d_{2})=0$, $g(d_{2},d_{1})=0$ to the problem of locating the intersection points in $\vartheta \in [0, \pi/4]$ of the two polar curves given by

\begin{small}
\begin{eqnarray}
R_{1}(\vartheta) &:=& {(4E+2) \sin \vartheta \sin 2\vartheta +{1\over 2} \sqrt{ (8E+4)^{2}\sin^{2}\vartheta \sin^{2}2\vartheta + 4\cos 2\vartheta \left( \sin 2\vartheta + \sin^{2}2\vartheta \right) }  \over \sin 2\vartheta + \sin^{2}2\vartheta } \nonumber \\
R_{2}(\vartheta) &:=& {(4E+2) \cos \vartheta \sin 2\vartheta +{1\over 2} \sqrt{ (8E+4)^{2}\cos^{2}\vartheta \sin^{2}2\vartheta - 4\cos 2\vartheta \left( \sin 2\vartheta + \sin^{2}2\vartheta \right) }  \over \sin 2\vartheta + \sin^{2}2\vartheta }
\end{eqnarray}
\end{small}

A solution of $R_{1}(\vartheta)=R_{2}(\vartheta)$ is given by $\vartheta = \pi/4$. This solution is the only critical point which  satisfies the stronger condition $\langle \varphi_{j} \vert a^{\dagger}a \vert \varphi_{j} \rangle = E$, $j=1,2$, because it is the only critical point for which the states $\ket{\varphi_{1}}$, $\ket{\varphi_{2}}$ have equal magnitudes of displacement and equal magnitudes of squeezing. In terms of the parameters appearing in Eq.(\ref{eqn:constr2}), the $\vartheta = \pi/4$ solution corresponds to  $d_{1}=d_{2}=2E+1$, $r_{2}  =  \sqrt{ {2E - f(d_{1},d_{2}) \over 2} }= \sqrt{ {E^{2}+E \over 2E+1} }$, (at this point, the determinant of the Hessian of Eq.(\ref{eqn:constr3}) is given by $e^{-8E^{2}-8E}(8E^{2}+8E+1)/2(2E+1)^{2} > 0$, so it is a minimum). For $E\rightarrow \infty$, $R_{1}(\vartheta) \rightarrow {8E\sin \vartheta \sin 2\vartheta \over \sin 2\vartheta + \sin^{2}2\vartheta}$ and $R_{2}(\vartheta) \rightarrow {8E\cos \vartheta \sin 2\vartheta \over \sin 2\vartheta + \sin^{2}2\vartheta}$; therefore, $\vartheta = \pi /4$ is the only solution in this limiting case. For finite $E$, Fig.\ref{fig:polar} shows one further solution in $\vartheta \in [0,\pi/4]$ (which has a reflected counterpart in the interval $(\pi/4 , \pi/2]$). This solution does not become degenerate with the $\vartheta = \pi/4$ solution for any $E>0$.

Note that a lower bound for the solution of the more strongly constrained problem, viz., the problem of minimizing $\vert \langle \varphi_{1} \vert \varphi_{2} \rangle \vert^{2}
$ with $\ket{\varphi_{j}}\in G(\mathcal{H})_{E}$, $j=1,2$, can be obtained from the smallest value of Eq.(\ref{eqn:constr3}) that also satisfies the stronger constraint. The solution of $R_{1}(\vartheta)=R_{2}(\vartheta)$ given by $\vartheta = \pi/4$ is a unique minimizer of Eq.(\ref{eqn:constr3}) that satisfies the stronger constraint; this point corresponds to the same pair of states  $\ket{\varphi_{1}}=\ket{  \left( - r(d_{c}(E)) , {1\over 2}\ln d_{c}(E) \right)}$, $\ket{\varphi_{2}}=\ket{  \left(  r(d_{c}(E)) , {1\over 2}\ln d_{c}(E) \right)}$ appearing in Lemmas 1 and 2. Therefore, $\min_{\ket{\varphi_{1}}, \ket{\varphi_{2}} \in G(\mathcal{H})_{E}} \vert \langle \varphi_{1} \vert \varphi_{2} \rangle \vert^{2} = \min_{\ket{\varphi_{1}}, \ket{\varphi_{2}} \in \mathcal{G}}\vert \langle \varphi_{1} \vert \varphi_{2} \rangle \vert^{2}=e^{-4E^{2}-4E}$. $\square$

Because the unitary rotation $U_{\theta}=e^{-i\theta a^{\dagger}a}$, $\theta \in  [0,2\pi)$, is the only element of the one-mode metaplectic representation of $Sp(2,\mathbb{R})$ that preserves the energy constraint, the set of pairs of energy constrained Gaussian states $(\ket{\varphi_{1}},\ket{\varphi_{2}})$ that achieve the minimal fidelity $\vert \langle \varphi_{1} \vert \varphi_{2} \rangle \vert^{2}=e^{-4E^{2}-4E}$ can be obtained from the single pair  $\ket{\varphi_{1}}=\ket{  \left( - r(d_{c}(E)) , {1\over 2}\ln d_{c}(E) \right)}$, $\ket{\varphi_{2}}=\ket{  \left(  r(d_{c}(E)) , {1\over 2}\ln d_{c}(E) \right)}$ by the application of $U_{\theta}$ to $\ket{\varphi_{1}}$ and $\ket{\varphi_{2}}$.

\section{Maximal trace distance between isoenergetic multimode bosonic Gaussian states\label{sec:multi}}

We now briefly consider the problem of maximization of the trace distance between two Gaussian states $\rho_{1}$, $\rho_{2} \in B_{1}(\mathcal{H}_{M})$, $\mathcal{H}_{M}:=\ell^{2}(\mathbb{C})^{\otimes M}$ for $M\in \mathbb{N}$, and subject to the isoenergetic constraint $\langle \sum_{j=1}^{M}a_{j}^{\dagger}a_{j} \rangle_{\rho_{k}} = ME$, where $k=1,2$ and $a_{j}$ ($a_{j}^{\dagger}$) is the annihilation (creation) operator on the $j$-th tensor factor.
Using Eq.(\ref{eqn:fidelrestate}) and the fact that the minimal fidelity occurs for two pure states, we state the problem symbolically as:

\underline{Problem (P2)}
\begin{eqnarray}
\ket{\varphi_{j}} \in G(\mathcal{H}_{M}) &\; , \;& j=1,2 \nonumber \\
 \vert \langle \varphi_{1} \vert \varphi_{2} \rangle \vert^{2} &\rightarrow& \mathrm{ min}, \nonumber \\
 \mathrm{tr}\left( \left(\sum_{j=1}^{M}a_{j}^{\dagger}a_{j}\right)\ket{\varphi_{k}}\bra{\varphi_{k}} \right)& = & ME \, , \, k=1,2  \, , \, M>1\nonumber .
\end{eqnarray}

In the special case that occurs when one restricts to pairs of symmetric Gaussian states of the form\\ $\left( \ket{\varphi_{1}}:=\ket{\omega}^{\otimes M} , \ket{\varphi_{2}}:= \ket{\sigma}^{\otimes M} \right)$, where $\ket{\omega}$, $\ket{\sigma} \in G(\mathcal{H}_{1})_{E}$, it is an immediate corollary of Theorem 1 that the minimal fidelity in this restricted optimization problem is given by $e^{-4ME^{2} - 4ME}$. However, much lower fidelity is achievable in the problem (P2), due to the fact that by taking all of the energy in a single tensor factor, i.e., taking, for example, \begin{eqnarray}\ket{\varphi_{1}}&=&\ket{0  }_{1} \otimes \ldots \otimes \ket{0}_{M-1}\otimes \ket{\left( - r(d_{c}(ME)) , {1\over 2}\ln d_{c}(ME) \right)}_{M} \nonumber \\ \ket{\varphi_{2}}&=&\ket{  0}_{1} \otimes \ldots \otimes \ket{0}_{M-1}\otimes \ket{\left(  r(d_{c}(ME)) , {1\over 2}\ln d_{c}(ME) \right)}_{M},\label{eqn:allin}\end{eqnarray} the solution of problem (P2) is at most $e^{-4M^{2}E^{2} - 4ME}$. In fact, if the $\ket{\varphi_{j}}$ appearing in problem (P2) are restricted to the submanifold of pure, separable states of $G(\mathcal{H}_{M})$, i.e., if the optimization is carried out over $\left( \ket{\varphi_{1}}:=\bigotimes_{j=1}^{M} \ket{\omega_{j}} , \ket{\varphi_{2}}:= \bigotimes_{j=1}^{M}\ket{\sigma_{j}} \right)$, where $\ket{\omega_{j}}$, $\ket{\sigma_{j}} \in G(\mathcal{H}_{1})_{E}$, then the solution of problem (P2) is exactly $e^{-4M^{2}E^{2} - 4ME}$. This result follows from the multiplicativity of the fidelity under tensor products.  

The presence of $M$ tensor factors in problem (P2) increases the degeneracy of pairs $(\ket{\varphi_{1}},\ket{\varphi_{2}})$ that exhibit maximal trace distance. Specifically, when unitarily represented on the Hilbert space $\mathcal{H}_{M}$ by the metaplectic representation, the maximal compact subgroup $O(2M,\mathbb{R})$ of $Sp(2M,\mathbb{R})$ is the maximal subgroup that preserves the total energy of $\ket{\varphi_{1}}$ and $\ket{\varphi_{2}}$ and their trace distance. One implication of this degeneracy is that the value $e^{-4M^{2}E^{2} - 4ME}$ for the fidelity can be obtained on pairs of pure states in the symmetric subspace of $\mathcal{H}_{M}$. To see this, simply consider the states in Eq.(\ref{eqn:allin}) after the transformation $a_{j}^{\dagger}\rightarrow {1\over \sqrt{M}}\sum_{\ell = 1}^{M}e^{-{2\pi i \ell j \over M} }a_{\ell}^{\dagger}$.

The main difficulty in Section \ref{sec:calcn} was to show that any maximally trace distant pair of isoenergetic, single mode bosonic Gaussian states is defined by a single covariance matrix and, therefore, that any such pair consists of opposite coherent states with respect to a certain complex structure on the symplectic space $(\mathbb{R}^{2},\Delta)$. We call two states $\rho_{1}$, $\rho_{2}\in G(\mathcal{H}_{M})$ \textit{isocovariant} if their respective covariance matrices are equal, i.e., if $\Sigma_{1}=\Sigma_{2}$. Lemma 3 simplifies the calculation of the maximal trace distance between two multimode, isoenergetic, isocovariant bosonic Gaussian states. The possibility that the solutions to problem (P2) are isocovariant pairs of multimode, isoenergetic bosonic Gaussian states can be motivated by a scaling argument, which we provide at the end of this section. 

\textbf{Lemma 3.} \textit{For $A\in M_{2M}(\mathbb{C})$, the $C^{*}$-algebra of $2M \times 2M$ complex matrices, let $\Vert A \Vert$ denote the operator norm on and} $\text{Tr}A$ \textit{the non-normalized matrix trace. Then,} \begin{eqnarray}
\min_{\substack{( \ket{\varphi_{1}}, \ket{\varphi_{2}} ) \in G(\mathcal{H}_{M})_{E} \\ \Sigma_{1}=\Sigma_{2}} } \vert  \langle \varphi_{1} \vert \varphi_{2} \rangle \vert^{2} &\ge& \min_{\substack{\det\Sigma = 4^{-M} \\ \text{Tr}\Sigma \le 2ME+M } }e^{-\Vert \Sigma^{-1} \Vert \left( 2ME +M - \text{Tr}\Sigma \right)} \nonumber \\ &=&  \min_{\substack{S \in Sp(2M,\mathbb{R}) \\  \text{Tr}S^{T}S \le (2E+1)\text{rank}(S^{T}S) } }e^{-\Vert S^{T}S \Vert \left( (2E+1) \text{rank}(S^{T}S) -\text{Tr}S^{T}S \right)},
\label{eqn:prop3}
\end{eqnarray}
\textit{where the constrained minimization on the right hand side of the first line is carried out over covariance matrices $\Sigma$, i.e., $\Sigma \in M_{2M}(\mathbb{R})_{+}$ and $\Sigma \ge \pm {i\over 2}\Delta$.}

\textbf{Proof.} The constraint in problem (P2) can be restated as $-{M\over 2}+{1\over 2}\text{Tr}\Sigma_{j} + {1\over 2}\Vert m_{j} \Vert^{2} = ME$, $j=1,2$. It follows that Eq.(\ref{eqn:purefidel}) can be rewritten as
\begin{eqnarray}
\vert \langle \varphi_{1} \vert \varphi_{2} \rangle \vert^{2} &\ge& {e^{-{1\over 2}\Vert (\Sigma_{1}+\Sigma_{2})^{-1} \Vert \, \Vert m_{1} - m_{2} \Vert^{2} } \over \sqrt{\text{det}\Sigma_{1} + \Sigma_{2} } } \nonumber \\ &\ge & {e^{-{1\over 2}\Vert (\Sigma_{1}+\Sigma_{2})^{-1} \Vert \, \left( \Vert m_{1} \Vert + \Vert m_{2} \Vert \right)^{2} } \over \sqrt{\text{det}\Sigma_{1} + \Sigma_{2} } } \nonumber \\ {}&=& {e^{-{1\over 2}\Vert (\Sigma_{1}+\Sigma_{2})^{-1} \Vert \, \left( \sum_{j=1}^{2}\sqrt{2ME+M-\text{Tr}\Sigma_{j}} \right)^{2} } \over \sqrt{\text{det}\Sigma_{1} + \Sigma_{2} } }.
\label{eqn:multimodefidel}
\end{eqnarray}
If $\Sigma_{1}=\Sigma_{2}$, then $\text{det} \Sigma_{1} + \Sigma_{2} =1$ follows immediately from Williamson's theorem (see \cite{holevoqubook}, Exercise 12.19). Furthermore, in this case, both mean vectors $m_{1}$ and $m_{2}$ lie on the sphere $S^{2M-1}$ of radius $\sqrt{2ME + M -\text{Tr}\Sigma_{1}}$, and $\Vert m_{1} - m_{2} \Vert$ is maximized for $m_{1}=-m_{2}$. By taking the minimum over both sides of (\ref{eqn:multimodefidel}) with $\Sigma_{1}=\Sigma_{2}$, the inequality appearing in the lemma is proven. Note that in this inequality, the constraints on $\Sigma$ correspond to the restriction to pure states with expected energy $2E$. Because $\Sigma$ corresponds to the covariance matrix of a pure Gaussian quantum state, there exists $S\in Sp(2M,\mathbb{R})$ such that $\Sigma = {1\over 2}S^{T}S$. In particular, $2\Sigma \in Sp(2M,\mathbb{R})$. Because the spectrum of a positive definite symplectic matrix $T$ takes the form $\bigcup_{j=1}^{{1\over 2}\text{rank}(T)}\lbrace \lambda_{j},\lambda_{j}^{-1} \rbrace$ \cite{gossonbook}, it follows that $\Vert \Sigma^{-1} \Vert = 2\Vert S^{T}S \Vert$, where we have noted that the number of modes, $M$, is equal to ${1\over 2}\text{rank}(\Sigma)$. The final line of Eq.(\ref{eqn:prop3}) can be further simplified, according to specific need. Note the the determinant constraint on $\Sigma$ (i.e., the restriction to pure Gaussian states) allows the representation in terms of symplectic matrices on in the final line of Eq.(\ref{eqn:prop3}); this follows from the fact that $\det S =1$ for all $S \in Sp(2M,\mathbb{R})$. $\square$

In fact, equality holds in the inequality of Lemma 3. By showing this, one arrives at an immediate solution of the problem of maximal trace distance between isoenergetic, isocovariant bosonic Gaussian states.

\textbf{Theorem 2.} $\displaystyle \min_{\substack{\ket{\varphi_{1}}, \ket{\varphi_{2}} \in G(\mathcal{H}_{M})_{E}\\ \Sigma_{1}=\Sigma_{2}}}\vert  \langle \varphi_{1} \vert \varphi_{2} \rangle \vert^{2} = e^{-4M^{2}E^{2}-4ME}$

\textbf{Proof.} The proof is a direct calculation that follows from Lemma 3. Taking $\lambda_{1}$ to be the maximal eigenvalue of $S^{T}S$ without loss of generality, the final line of Eq.(\ref{eqn:prop3}) can be written \begin{equation}
\min_{\substack{( \ket{\varphi_{1}}, \ket{\varphi_{2}} ) \in G(\mathcal{H}_{M})_{E} \\ \Sigma_{1}=\Sigma_{2}} } \vert  \langle \varphi_{1} \vert \varphi_{2} \rangle \vert^{2}\ge \min_{\sum_{j=1}^{M}\lambda_{j}+\lambda_{j}^{-1} \le 4ME+2M  }e^{-2\lambda_{1} \left( 2ME +M - {1\over 2}\sum_{j=1}^{M}\lambda_{j}+\lambda_{j}^{-1} \right)}.
\label{eqn:simplifybound}\end{equation} Taking derivatives of the exponential function on the right hand side with respect to $\lambda_{j}$ gives the critical values $\lambda_{1}=2ME+1$ and $\lambda_{j}=1$ for $j=2,\ldots , M$, which satisfy the energy constraint. The corresponding minimal fidelity is $e^{-4M^{2}E^{2}-4ME}$. This value of the fidelity can be achieved for the pair $(\ket{\varphi_{1}},\ket{\varphi_{2}})$ in Eq.(\ref{eqn:allin}); therefore, the bound in Eq.(\ref{eqn:simplifybound}) is an equality. $\square$

Due to Eq.(\ref{eqn:allin}) and Theorem 2, it follows that the maximal isocovariant trace distance can be obtained on pairs of states with maximal energy density, i.e., pairs of states that have single mode marginal states that satisfy the energy constraint. However, the full solution of the problem of finding maximal trace distance between isoenergetic bosonic Gaussian states of $M$ modes requires that one perform the maximization of the trace distance over all pairs $(\rho_{1},\rho_{2}) \in G(\mathcal{H}_{M})^{\times 2}$. Note that in the single mode case, the unconstrained optimization over $\theta_{j}$ in Lemma 2 allowed to reduce the problem in Theorem 1 to the case of commuting covariance matrices, i.e., to the case of $[\Sigma_{1},\Sigma_{2}]=0$. However, for the multimode case ($M>1$), even if one loosens the restriction of isocovariance to consider the case of commuting covariance matrices, the constrained optimization problem over the spectra of $\Sigma_{1}$ and $\Sigma_{2}$ is considerably more challenging than the problem posed by Theorem 1. The resulting expressions may be amenable to numerical optimization for small, finite $M$.

There are qualitative considerations that suggest that the minimum value appearing in Theorem 2 indeed coincides with the solution of problem (P2) for any $M$. For example, the determinant appearing in the denominator of Eq.(\ref{eqn:purefidel}) contributes only $\mathcal{O}(\log E)$ to the exponent of the minimal fidelity, whereas the quadratic form appearing in the exponent scales polynomially with $E$. This suggests that to reduce the fidelity between two isoenergetic Gaussian states, the pair of states should be taken isocovariant. Furthermore, given $M>1$, it is also true that if the minimum value appearing in Theorem 2 were greater than the minimum in problem (P2), then it would follow from Theorem 1 that an energy value greater than $ME$ is required for a pair of isoenergetic, single mode bosonic Gaussian states to achieve the minimal fidelity that can be obtained in the case of two $M$-mode, isoenergetic bosonic Gaussian states.


\section{Applications}

The main results of sections \ref{sec:calcn} and \ref{sec:multi} have several applications to the theory of low power quantum communication. In particular, a central problem in optical detection theory consists of the construction of optimal quantum measurements that saturate the minimal error probability in the task of discrimination of a finite set of pure quantum signals \cite{helstrombook,holevoasymptotic}. An extension of this problem beyond the traditional settings of, e.g., narrow band coherent signals, or signals generated by the action of a finite cyclic unitary group, was provided in Ref.\cite{barnettdiscrim}. The present work motivates the investigation of optimal discrimination of a finite set of multimode, isoenergetic pure Gaussians which is generated from a single Gaussian state by a discrete subgroup of the energy conserving (i.e., compact) subgroup of $Sp(2M,\mathbb{R})$.  In the study of asymptotic quantum hypothesis testing in the independent, identically distributed (i.i.d.) setting \cite{amarinagaoka}, the quantum Pinsker inequality allows one to bound the quantum relative entropy by a multiple of the squared trace distance, thereby relating the error of the second kind \cite{ogawanagaoka} to the minimal error probability for distinguishing two quantum states. The recent calculation of the Petz-R\'{e}nyi relative entropies of Gaussian states \cite{wilderenyi} provides further motivation for the use of Gaussian states as model systems for asymptotic quantum hypothesis testing.

On the mathematical side, given a (dual) completely positive, unital map $\Phi^{*}: B(\ell^{2}(\mathbb{C}))    \rightarrow B(\mathcal{K}) $ such that $\mathcal{K}$ is a Hilbert space of at most countably infinite dimension and such that the image of $\Phi$ is contained in the set of Gaussian states, one can define an energy-constrained Gaussian trace norm contraction coefficient as $\eta_{\mathcal{S}}:= \sup_{\rho_{1},\rho_{2}\in \mathcal{S}} {\Vert \Phi(\rho_{1})-\Phi(\rho_{2}) \Vert_{1}\over \Vert \rho_{1}-\rho_{2}\Vert_{1} }$, where $\mathcal{S}$ is the set of (densities of) normal states on $B(\mathcal{K})$ (see Ref.\cite{ruskaiimproved,ruskaicontract} for analyses of contraction coefficients and applications). Knowledge of the maximal trace distance in $G(\ell^{2}(\mathbb{C}))_{E}$ for any $E$ allows one to bound $\eta_{\mathcal{S}}$. Such completely positive, unital maps appear in the analysis of dissipative dynamics of the electromagnetic field coupled to atomic matter. Furthermore, the structure of linear bosonic, non-Gaussian quantum channels defined by Stinespring isometries that map Fock states $\ket{n}$, $n\in 0,1,\ldots$, to $V\ket{n} \otimes (\ket{  \left( - r(d_{c}(E)) , {1\over 2}\ln d_{c}(E) \right) } + \ket{ \left(  r(d_{c}(E)) , {1\over 2}\ln d_{c}(E) \right) })/\mathcal{N} $, where $V$ is a metaplectic representation of an element of $Sp(4,\mathbb{R})$ and $\mathcal{N}$ is a normalization constant, have been explored in the context of generation of entanglement and nonclassicality \cite{volklinbosonic}.  Finally, we note that the resource theory of quantum coherence, which has recently been extended to quantum optics \cite{tanvolk}, provides a rigorous framework for the conversion rates of Gaussian and non-Gaussian states under a physically-motivated class of quantum dynamics. This progress allows further exploration of the uses of maximally distant pairs of constrained Gaussian states in quantum information processing.

\section{Conclusion}

In this work, we have derived the pairs of Gaussian states $(\rho_{1},\rho_{2})$ of a single quantum harmonic oscillator that exhibit maximal trace distance $ \Vert \rho_{1} - \rho_{2} \Vert_{1}$ subject to the energy constraint $\langle a^{\dagger}a \rangle_{\rho_{1}} = \langle a^{\dagger}a \rangle_{\rho_{2}} = E $; in particular, we have derived the corresponding mean vectors and covariance matrices of these pairs. We have also derived the set of maximally trace distant isoenergetic, isocovariant Gaussian states of the multimode quantum electromagnetic field. In both cases, each optimal pair consists of nonclassical pure states which are related by a local rotation in phase space. Given an optimal pair, the full set of optimal pairs can be constructed by the action of the symplectic subgroup that preserves the energy constraint.

We conclude by mentioning two major challenges for the theory of minimal error distinguishability of Gaussian states: 1) a general formula for the trace distance between Gaussian states, and 2) calculation of the maximal trace distance between isoenergetic bosonic Gaussian states given constraints not only on total energy, but on, e.g., physically-motivated quantities such as energy distribution, entropy, entanglement, and squeezing \cite{wolfsqueez}.  Furthermore, it may be fruitful to obtain shorter proofs of Theorems 1 and 2 in terms of dynamics, e.g., based on the properties of linear bosonic Gaussian quantum channels and their information-theoretic characterizations, to contrast with the ``kinematical'' proofs provided here. In general, the results of the present work provide a basis for future studies in the theory of optimal detection of Gaussian continuous variable quantum signals under local energy constraints.

\acknowledgements

The author thanks Y. Kwon for hosting during the completion of this work and A.S. Holevo for useful comments. This work was supported by the Korea Research Fellowship Program through the National Research Foundation of Korea (NRF) funded by the Ministry of Science and ICT (2016H1D3A1908876) and by the Basic Science Research Program through the National Research Foundation of Korea (NRF) funded by the Ministry of Education (2015R1D1A1A09056745).

\section{Appendix}

In the proof of Theorem 1, we stated without proof the fact that centered, pure Gaussian states $\ket{\varphi_{1}}$, $\ket{\varphi_{2}}$ with energy $2E$ distributed between them, satisfy $\vert \langle \varphi_{1} \vert \varphi_{2} \rangle \vert^{2} > e^{-4E^{2}-4E}$ for all $E>0$. This is a simple consequence of the fact that all the energy $2E$ must go into squeezing the states. Let $\ket{\varphi_{1}}= S(w_{j})\ket{0}$ with $w_{j} \in \mathbb{R}$. Then
\begin{equation}
\vert \langle 0 \vert S(w_{1})^{\dagger}S(w_{2}) \vert 0 \rangle \vert^{2} = {1\over \cosh (w_{1}-w_{2})}.
\end{equation}
Using $\sinh^{2}w_{1}+\sinh^{2}w_{2} = 2E$ to eliminate $w_{2}$ gives 
\begin{equation}
\vert \langle 0 \vert S(w_{1})^{\dagger}S(w_{2}) \vert 0 \rangle \vert^{2} =  {1\over \cosh (\sinh^{-1}\sqrt{2E - \sinh^{2}w_{1}} - w_{1})}.
\end{equation}
The minimum fidelity $\vert \langle 0 \vert S(w_{1})^{\dagger}S(w_{2}) \vert 0 \rangle \vert^{2} = 1/(2E+1) \ge e^{-4E^{2}-4E}$ occurs at $w_{1}=-\sinh^{-1}\sqrt{E}$, $w_{2}=-w_{1}$.

\bibliography{catbib.bib}

\end{document}